\documentstyle[11pt,paspconf,epsf]{article}

\begin{document}

\title{Deep STIS Luminosity Functions for LMC Clusters}

\author{R. Elson, N. Tanvir, G. Gilmore, R.A. Johnson, \& S. Beaulieu} 

\affil{Institute of Astronomy, Madingley Rd., Cambridge CB3 0HA, England
email: elson,nrt,gil,raj,beaulieu @ast.cam.ac.uk}

\begin{abstract}
We present deep luminosity functions derived from HST STIS data for
three rich LMC clusters (NGC 1805, NGC 1868, and NGC 2209), 
and for one Galactic globular cluster (NGC 6553).
All of the LMC cluster luminosity functions are roughly consistent with a
Salpeter IMF or with the solar neighbourhood IMF from Kroupa, Tout
\& Gilmore (1993).  They continue
to rise at least  to 0.7$M_\odot$.  NGC 1868
shows evidence for mass segregation which may be primordial.  A
comparison of deep luminsoisty functions for seven Galactic globulars
shows that the luminosity functions are eroded at low masses by
amounts that are strongly correlated with distance 
from the Galactic plane.  

\end{abstract}


\section{Introduction and Data}

The LMC is an ideal laboratory for studying the formation and evolution
of rich star clusters.  We describe the first results from a 
95-orbit Cycle 7 HST project (No. 7307) with this as its primary aim
(Elson et al. 1997).
Briefly, we are using WFPC2, NICMOS2
and STIS (in imaging mode with the F28X50LP filter) 
to obtain deep $(V-H)$ CMDs and $\sim R$-band
luminosity functions for eight clusters:
NGC 1805 and NGC 1818 ($\sim 10^7$ yr), NGC 1831 and NGC 1868 
($\sim 10^8$ yr),
NGC 2209 and Hodge 14 ($\sim 10^9$ yr), and NGC 2210 and Hodge 11 
($\sim 10^{10}$ yr).
We also have data for the Galactic globular cluster NGC 6553, 
primarily for calibration purposes.

These data will allow us to determine age spreads in the young clusters
and identify pre-main-sequence stars (and thus investigate the timescale
and sequence of star formation), to quantify the binary population
and trace its evolution and the development of mass segregation.  We 
will also be able to investigate the universality of the IMF, which is
the focus of this presentation.  NGC 1868 and NGC 2209 are particularly
interesting in this regard.  They have similar ages but very different
core radii, and one possibility is that different IMFs ($x\sim 1$ as
opposed to $x \sim 2$, where $x=1.35$ is the Salpeter value)
have caused their cores to expand (through mass loss due to stellar
evolution) at different rates (Elson et al. 1989).

We determined magnitudes using PSF fitting, with  PSFs constructed
from stars in the images.  Completeness was
determined using artificial star tests,  and background
contamination using STIS images of a field near each
cluster. 
STIS magnitudes were transformed to absolute magnitudes in the HST system using
a zero-point of $K_{STIS}=23.4$ (H. Ferguson, 1998, private communication),
a distance modulus of 18.5, and an absorption $A_{STIS}=0.18$.

\section{Results and Discussion}

Figure 1 shows  luminosity functions for NGC 1805, NGC 1868 and  
NGC 2209.   The LFs are for stars near the half-mass radius where
any dynamical evolution should not affect the mass function. 
All three are similar.
Also shown are LFs corresponding to  
power-law IMFs with slopes $x=1.0,1.5,2.0$. 
All the luminosity functions continue to rise at masses $< 0.7 M_\odot$.  
Variations in IMF slope do not appear to be responsible for the
differences in core radii of NGC 1868 and NGC 2209.
The LFs appear to be similar to those of Galactic globular clusters.
The rising LF at low masses suggests that the clusters will survive
disruption through evaporation, and
evolve to objects like the old LMC clusters.  They also suggest that
30 Doradus, which is probably a younger counterpart of these clusters,
will reveal a similar population of low mass stars (cf. H. Zinnecker, this 
volume).  This is important, as 30 Dor has often been cited as an
archetypal starburst, assumed to have an IMF truncated at relatively
high mass.

\begin{figure}
\vspace{0.0in}
\centerline{\plotfiddle{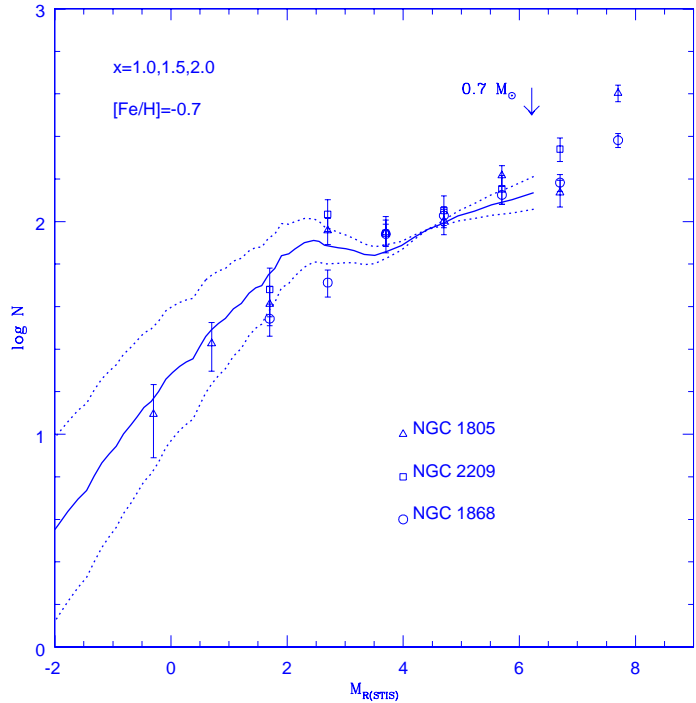}{7.0cm}{0}{50}{50}{-365}{-80}}
\caption{Luminosity functions for three LMC clusters compared to model
power-law IMFs with $x=1.0, 1.5, 2.0$. Arbitrary 
shifts in $\log N$ have been applied.} 
\label{fig-1}
\end{figure}

Figure 2 shows luminosity functions for NGC 1868 derived at three
different radii.  There is clear evidence for mass segregation. 
N-body modelling, which is an integral part of our
project, will indicate to what extent this is primordial
or due to dynamical evolution.

\begin{figure}
\vspace{0.0in}
\centerline{\plotfiddle{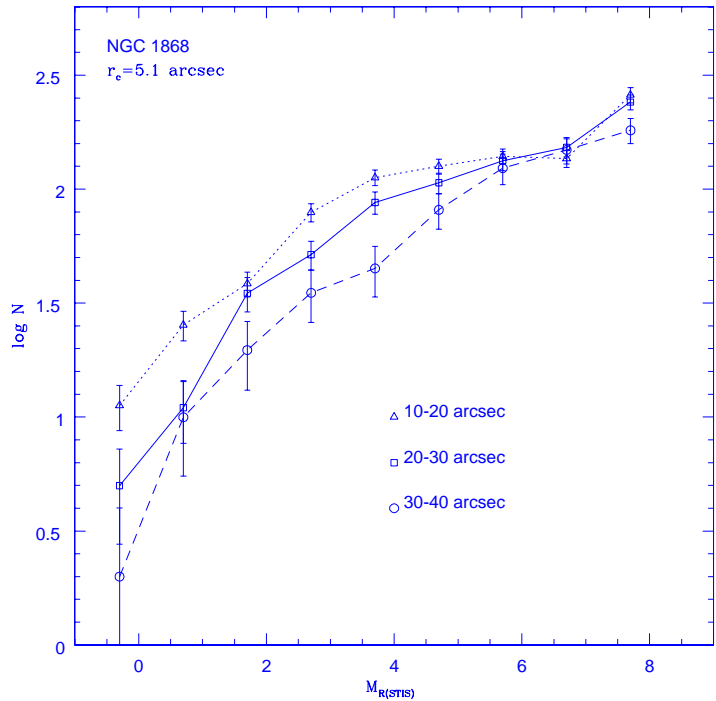}{7.0cm}{0}{50}{50}{-365}{-80}}
\caption{Luminosity functions for NGC 1868 at three different radii.} 
\label{fig-2}
\end{figure}

Figure 3 shows the LF for NGC 6553 and those of six other
Galactic globulars from the literature (Elson et al 1995; Santiago et al
1996; Piotto et al 1997).  A transformation from STIS magnitude to $M_{814}$
was derived from the Padova isochrones (G. Worthey, 1998, private
communication).  While the luminosity functions
agree well at brighter magnitudes, the faint ends differ markedly.
Such differences have been noted before,
and have been attributed to either metallicity effects, evaporation of 
low mass stars,
or stripping of low mass stars by tidal shocking (cf. Piotto et al 1997).
We quantified
the difference among the luminosity functions as the increment $\Delta\log
N$ at $M_{814}=8.0,8.5,9.0$. 
Figure 4 shows $\Delta\log N$ plotted against the distance, $Z$, of
each cluster from the Galactic plane.  There is a striking correlation,
which suggests that tidal
shocking is primarily responsible for the differences among  the
luminosity functions.   A plot of $\Delta\log N$ against metallicity
($-0.3 < $ [Fe/H] $ < -2.5$ for this sample) shows no correlation.

\begin{figure}
\vspace{0.0in}
\centerline{\plotfiddle{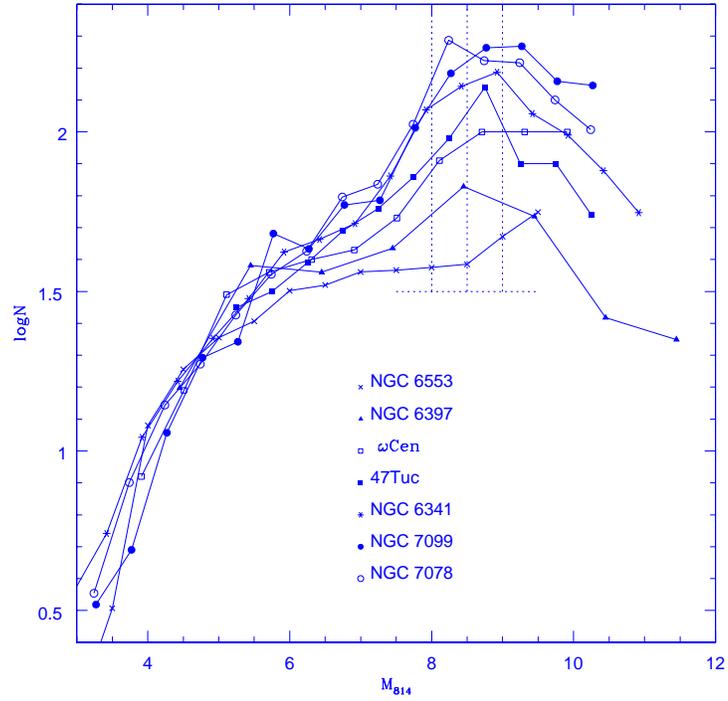}{7.0cm}{0}{50}{50}{-365}{-80}}
\caption{Luminosity functions for 7 Galactic globulars.} 
\label{fig-3}
\end{figure}

\begin{figure}
\vspace{0.0in}
\centerline{\plotfiddle{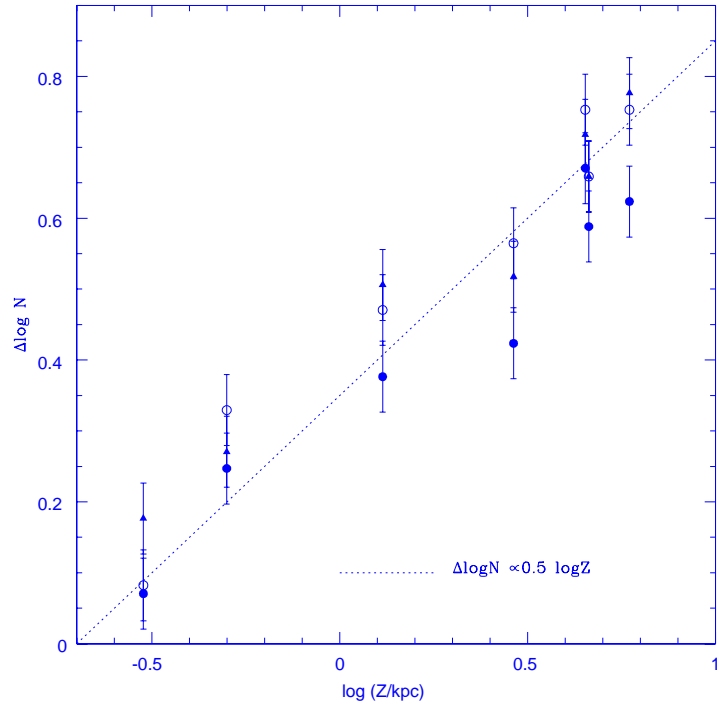}{8.0cm}{0}{50}{50}{-365}{-80}}
\caption{Relative flattening of the faint end of the
luminosity functions in Fig. 3, measured
at $M_{814}=8.0,8.5,9.0$ (filled circles, open circles, triangles)
plotted against
distance above the Galactic plane.} 
\label{fig-4}
\end{figure}

We are awaiting STIS images of four more clusters, NGC 1818, NGC 1831,
Hodge 14, and Hodge 11.  Also, background data still to be acquired , as
well as improved data reduction methods, will allow us to push our
luminosity functions $1 - 2$ mag fainter than at present (and to 
verify the upturn at $M_{814} > 8.5$ in the LF of NGC 6553).

\end{document}